\documentstyle[12pt]{article}

\newcommand{\be}{\begin{equation}}
\newcommand{\ee}{\end{equation}}
\newcommand{\bea}{\begin{eqnarray}}
\newcommand{\eea}{\end{eqnarray}}

\newcommand{\Int}{\int\limits}
\newcommand{\ep}{\varepsilon}
\newcommand{\Li}[2]{{\mbox{Li}}_{#1}\left(#2\right)}

\newcommand{\Cl}[2]{{\mbox{Cl}}_{#1}\left(#2\right)}
\newcommand{\Ls}[2]{{\mbox{Ls}}_{#1}\left(#2\right)}
\newcommand{\Ti}[2]{{\mbox{Ti}}_{#1}\left(#2\right)}

\def\thebibliography#1{\centerline{\bf References}
\list
 {[\arabic{enumi}]}{\settowidth\labelwidth{[#1]}\leftmargin\labelwidth
 \advance\leftmargin\labelsep
 \usecounter{enumi}}
 \def\newblock{\hskip .11em plus .33em minus .07em}
 \sloppy\clubpenalty4000\widowpenalty4000
 \sfcode`\.=1000\relax}

\newcommand{\bi}[1]{\vspace{-2mm} \bibitem{#1}}

\begin{document}

\thispagestyle{empty}
 \begin{flushright}
 {\tt University of Bergen, Department of Physics}    \\[2mm]
 {\tt Scientific/Technical Report No.1995-07}    \\[2mm]
 {\tt ISSN 0803-2696} \\[5mm]
 {MZ-TH--95-14}\\[5mm]
 {hep-ph/9504431} \\[5mm]
 {April 1995}           \\
\end{flushright}
 \vspace*{3cm}
\begin{center}
 {\bf \large
A magic connection between massive and massless diagrams}
\vspace{10 mm} \\
  A.I.~Davydychev$^{a,}$
  \footnote{On leave from
  Institute for Nuclear Physics,
  Moscow State University, 119899 Moscow, Russia.\\
  E-mail address: davyd@vsfys1.fi.uib.no}
  and J.B.~Tausk$^{b,}$
  \footnote{E-mail address: tausk@vipmzw.physik.uni-mainz.de}
\vspace{10 mm} \\
$^{a}${\em
  Department of Physics, University of Bergen,\\
  All\'{e}gaten 55, N-5007 Bergen, Norway}
\vspace{3 mm} \\
$^{b}${\em
  Institut f\"ur Physik,
  Johannes Gutenberg Universit\"at,\\
  Staudinger Weg 7, D-55099 Mainz, Germany}
\end{center}
\vspace{12 mm}
\begin{abstract}
A useful connection between two-loop massive vacuum integrals
and one-loop off-shell triangle diagrams with massless
internal particles is established for arbitrary values
of the space-time dimension $n$.
\end{abstract}
\newpage
\setcounter{page}{2}
\setcounter{footnote}{0}

\newpage

{\bf 1.}
Many experiments testing the Standard Model and its extensions are
sensitive to the values of two-loop
corrections to physical quantities and require theoretical
predictions for these contributions. Since evaluating massive
two-loop diagrams is a tricky business, looking for non-trivial
connections between different
diagrams may be of certain interest.

 In a previous paper \cite{DT}, we noticed that, apart from
some simple logarithmic terms, the finite part
of the two-loop vacuum integral with three massive propagators
involves a non-trivial function of the masses, which is exactly
the same as a one-loop triangle with three massless propagators
whose external momenta squared are equal to those masses squared.
This suggests that there is a connection between these two seemingly
unrelated diagrams. The purpose of this paper is to provide an
explanation of this connection (for arbitrary values of the space-time
dimension) and to explore some of its consequences.

 One of the important applications of the results for massive two-loop
vacuum integrals is the small momentum expansion of two-loop diagrams
with non-zero
external momenta (see, e.g., in \cite{DT,FT}). Moreover, application
of the general theory of asymptotic expansions \cite{as-ex} shows that
analogous integrals also appear in the large momentum expansion
and in the ``zero-threshold'' expansion
(see in refs. \cite{(B)DST,LvRV-NPB}).
Another application is the calculation of two-loop contributions to
the $\rho$-parameter (see in \cite{Bij,2-loop-rho}).
Furthermore, three-loop calculations in dimensional regularization
\cite{dimreg} (with the space-time dimension $n~=~4~-~2\ep$)
require knowledge of the order $\ep$ contribution to two-loop vacuum
diagrams. As an example, we can mention recent calculations
of three-loop corrections to the $\rho$-parameter \cite{3-loop-rho},
and some other developments \cite{alpha3,LvRV-NPB}.

The present paper is organized as follows. In section~2 we
give definitions and different representations of the integrals
considered. In section~3 we present some results for dimensionally
regulated massless triangle diagrams. In section~4 we derive
the connection between massive and massless integrals and use
the corresponding results of section~3 to get the $\ep$-part
of two-loop vacuum integrals with different masses.
In section~5 we consider what the general results yield for
the important case of equal masses. In section~6 (conclusion)
we discuss the main results of the paper.

\newcommand{\partaa}
{\setlength{\unitlength}{1mm}
 \begin{picture}(50,28)(0,-16)
 \put(25,0){\circle{13}}
 \put(18.5,0){\circle*{1.0}}
 \put(31.5,0){\circle*{1.0}}
 \put(18.5,0){\line(1,0){13}}
 \put(25,8.0){\makebox(0,0)[c]{${\scriptstyle m_1, \nu_1}$}}
 \put(25,1.5){\makebox(0,0)[c]{${\scriptstyle m_2, \nu_2}$}}
 \put(25,-8.0){\makebox(0,0)[c]{${\scriptstyle m_3, \nu_3}$}}
 \put(25,-15){\makebox(0,0)[c]{(a)}}
 \end{picture}}
\newcommand{\partbb}
{\setlength{\unitlength}{1mm}
 \begin{picture}(50,28)(0,-16)
 \put(13,0){\line(1,0){8}}
 \put(21,0){\line(1,1){8}}
 \put(21,0){\line(1,-1){8}}
 \put(29,8){\line(1,0){8}}
 \put(29,-8){\line(1,0){8}}
 \put(29,-8){\line(0,1){16}}
 \put(21,0){\circle*{1.0}}
 \put(29,8){\circle*{1.0}}
 \put(29,-8){\circle*{1.0}}
 \put(31.5,0.0){\makebox(0,0)[c]{${\scriptstyle \nu_3}$}}
 \put(24,-5.5){\makebox(0,0)[c]{${\scriptstyle \nu_2}$}}
 \put(24,5.5){\makebox(0,0)[c]{${\scriptstyle \nu_1}$}}
 \put(17,1.5){\makebox(0,0)[c]{${\scriptstyle p_3}$}}
 \put(32,10.5){\makebox(0,0)[c]{${\scriptstyle p_2}$}}
 \put(32,-10.5){\makebox(0,0)[c]{${\scriptstyle p_1}$}}
 \put(25,-15){\makebox(0,0)[c]{(b)}}
 \end{picture}}

 \begin{figure}[b]
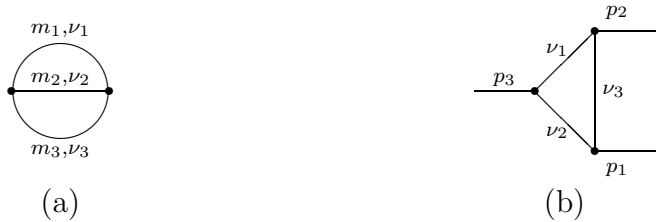

 \[
 \begin{array}{ccc}
 \partaa & $\hspace{1cm}$ & \partbb
 \end{array}
 \]
 \caption[]{Two-loop vacuum diagram (a) and one-loop triangle (b)}
 \label{figIJ}
 \end{figure}

\vspace{4mm}

{\bf 2.}
We shall discuss the following two types of
Feynman integrals (see Fig.~1a,b):
\be
\label{defI}
I(n; \; \nu_1, \nu_2, \nu_3 | m_1^2, m_2^2, m_3^2)
\equiv  \int \int \frac{\mbox{d}^n p \;\; \mbox{d}^n q}
      {\left( p^2 - m_1^2 \right)^{\nu_1}
       \left( q^2 - m_2^2 \right)^{\nu_2}
       \left( (p-q)^2 - m_3^2 \right)^{\nu_3}} ,
\ee
\be
\label{defJ}
J (n; \; \nu_1  ,\nu_2  ,\nu_3 | p_1^2, p_2^2, p_3^2) \equiv \int
 \frac{\mbox{d}^n r}{ ((p_2 -r )^2)^{\nu_1}  ((p_1 +r )^2)^{\nu_2}
      (r^2)^{\nu_3} } ,
\ee
where $p_3= -(p_1+p_2)$.
To establish a connection between (\ref{defI}) and (\ref{defJ}),
we shall consider $p_i^2=m_i^2$, and we shall use dimensionless
variables
\be
\label{xy}
x \equiv p_1^2/p_3^2 = m_1^2/m_3^2 \;\; , \;\;
y \equiv p_2^2/p_3^2 = m_2^2/m_3^2 .
\ee
Below we shall omit the arguments $m_i^2$
and $p_i^2$ in the integrals $I$ and $J$, respectively.

{\em Feynman parametric representations}:
\be
\label{FpI}
I(n; \nu_1, \nu_2, \nu_3) = \mbox{i}^{2-2\Sigma\nu_i} \; \pi^n \;
\frac{\Gamma(\sum \nu_i -\! n)}
     {\prod \Gamma(\nu_i)}
\int\!\!\!\!{\Int_{\alpha_i \geq 0}^{}}\!\!\!\!\int
\frac{\prod \alpha_i^{\nu_i -1}  \; \mbox{d} \alpha_i \;
            \delta\left(\sum\alpha_i -1 \right)}
{(\alpha_1 \alpha_2 \!+\! \alpha_1 \alpha_3 \!+\! \alpha_2 \alpha_3)^{n/2}
    \left( \sum \alpha_i m_i^2 \right)^{\Sigma\nu_i -n}} ,
\ee
\be
\label{FpJ}
J(n; \nu_1, \nu_2, \nu_3) = \mbox{i}^{1-n} \; \pi^{n/2} \;
\frac{\Gamma(\sum \nu_i - n/2)}
     {\prod \Gamma(\nu_i)}
\int\!\!\!\!{\Int_{\alpha_i \geq 0}^{}}\!\!\!\!\int
\frac{\prod \alpha_i^{\nu_i -1}  \; \mbox{d} \alpha_i \;
            \delta\left(\sum\alpha_i -1 \right)}
     {(\alpha_1 \alpha_2 p_3^2 + \alpha_1 \alpha_3 p_2^2
               + \alpha_2 \alpha_3 p_1^2)^{\Sigma\nu_i -n/2}} ,
\ee
where $\sum$ and $\prod$ denote the sum and the product from $i=1$ to 3,
respectively.
Because of the delta functions, the integrations in (\ref{FpI}) and
(\ref{FpJ}) can be restricted to $\alpha_i \leq 1$.
Nevertheless, we prefer to write them in the present form, to
simplify the discussion of parameter transformations
we are going to describe below.

{\em Mellin--Barnes contour integral representations}:
\bea
\label{MBI}
I(n; \nu_1, \nu_2, \nu_3)
= \frac{\pi^n \mbox{i}^{2-2 \Sigma\nu_i} (m_3^2)^{n- \Sigma\nu_i}}
       {\Gamma(n/2) \; \prod \Gamma(\nu_{i})}
\; \frac{1}{(2 \pi \mbox{i})^2}
\Int_{-\mbox{i}\infty}^{\mbox{i}\infty}
\Int_{-\mbox{i}\infty}^{\mbox{i}\infty}
\mbox{d}s\; \mbox{d}t\;
x^s \; y^t \;
\Gamma(-s) \Gamma(-t)
\hspace{16mm}
\nonumber \\
\times
\Gamma(n/2\!-\!\nu_1\!-\!s) \Gamma(n/2\!-\!\nu_2\!-\!t)
\Gamma(\nu_1+ \nu_2 -n/2 +s+t)
\Gamma\left(\sum\nu_i -n+s+t\right) \; ,
\eea
\bea
\label{MBJ}
J(n; \; \nu_1, \nu_2, \nu_3) =
\frac{\pi^{n/2} \; \mbox{i}^{1-n} \; (p_3^2)^{n/2- \Sigma \nu_i}}
  {\Gamma \left( n- \sum \nu_{i} \right) \; \prod \Gamma(\nu_{i})}
\; \frac{1}{(2\pi \mbox{i})^2}
\Int_{-\mbox{i}\infty}^{\mbox{i}\infty}
\Int_{-\mbox{i}\infty}^{\mbox{i}\infty}
\mbox{d}s \; \mbox{d}t \;
x^s \; y^t \;
\Gamma (-s) \Gamma (-t)
\hspace{10mm}
\nonumber \\
\times  \Gamma(n/2 - \nu_2 - \nu_3  -s)
\Gamma(n/2 - \nu_1  -  \nu_3  -t)
\Gamma(\nu_3 + s + t)
\Gamma\left(\sum\nu_i - n/2 + s + t\right) ,
\eea
where the integration  contours  are chosen so as to separate
the ``right'' and ``left'' series of poles of gamma functions in the
integrand.
By use of the residue theorem, the result for
arbitrary $n$ and $\nu_i$ can be found in terms of
hypergeometric functions of two variables
(see \cite{BD,JPA,DT}).

{\em Uniqueness condition}:
When the $\nu_i$ and $n$ are related by
$\nu_1+ \nu_2+ \nu_3 = n$, a very simple result can be obtained
from (\ref{FpJ}) for such a ``unique'' triangle \cite{uniq}
(see also in \cite{Gracey}):
\be
\label{uniq}
\left. \frac{}{}
J(n; \; \nu_1,\nu_2,\nu_3) \right|_{\Sigma \nu_i = n}
=  \pi^{n/2} \; \mbox{i}^{1-n} \prod_{i=1}^{3}
\frac{\Gamma ( n/2 - \nu_i)}{\Gamma (\nu_i)} \;
\frac{1}{(p_i^2)^{n/2 - \nu_i}}.
\ee

\vspace{4mm}

{\bf 3.}
In the paper \cite{UD3}, the following representation valid
for arbitrary $\ep$ was obtained:
\be
\label{int-dr}
J(4-2\ep;1,1,1) = \frac{\pi^{2-\ep}\;\mbox{i}^{1+2\ep}}{(p_3^2)^{1+\ep}} \;
                 \frac{\Gamma (1+\ep) \Gamma^2 (1-\ep)}{\Gamma(1-2\ep)} \;
                 \frac{1}{\ep} \Int_0^1
 \frac{\mbox{d}\xi\;\xi^{-\ep} \left( (y\xi)^{-\ep} - (x/\xi)^{-\ep} \right)}
      {\left( y \xi^2 + (1-x-y)\xi + x \right)^{1-\ep}} .
\ee
We managed to generalize this result to the case when one of the $\nu$'s
is arbitrary, as
\bea
\label{secret}
J(4-2\ep;1,1,1+\delta)
= \frac{\pi^{2-\ep}\;\mbox{i}^{1+2\ep}}{(p_3^2)^{1+\ep+\delta}} \;
  \frac{\Gamma(1-\ep) \; \Gamma(1-\ep-\delta) \; \Gamma(1+\ep+\delta)}
       {\Gamma(1+\delta) \; \Gamma(1-2\ep-\delta)} \;
\hspace{18mm}
\nonumber \\[1mm]
\times
 \frac{1}{\ep+\delta} \;
 \Int_0^1
 \frac{\mbox{d}\xi\;\xi^{-\ep}
 \left( (y\xi)^{-\ep-\delta} - (x/\xi)^{-\ep-\delta} \right)}
      {\left( y \xi^2 + (1-x-y)\xi + x \right)^{1-\ep}} .
\eea
At $\delta=0$, eq.~(\ref{secret}) gives (\ref{int-dr}), whilst for
$\ep=0$ we get eq.~(15) of \cite{UD3} (see also in \cite{UD1-2-4}).

The expansion of $J(4-2\ep;1,1,1)$ in $\ep$ is
\be
\label{J-ep}
J(4-2\ep;1,1,1) =
\pi^{2-\ep}\;\mbox{i}^{1+2\ep} (p_3^2)^{-1-\ep} \;
\Gamma (1+\ep) \;
\left\{ \Phi^{(1)}(x,y) + \ep \; \Psi^{(1)}(x,y) + {\cal O}(\ep^2)
\right\} ,
\ee
with
\be
\label{Phiint}
\Phi^{(1)}(x,y) = - \Int_0^1
\frac{\mbox{d} \xi}{y \xi^2 + (1-x-y) \xi +x} \;
\left( \ln\frac{y}{x} + 2\ln{\xi}  \right),
\ee
\be
\label{Psiint}
\Psi^{(1)}(x,y) = - \!\Int_0^1 \!\frac{\mbox{d} \xi}
{y \xi^2 \!+ \!(1 \! - \! x \! - \! y)\xi \!+\! x}
\left( \ln\frac{y}{x} \!+\! 2\ln\xi \right) \!
\left[ \ln\left(\!\frac{y \xi^2 \!+\! (1\! - \! x \! - \! y)\xi \!+\! x}{\xi}
     \! \right)
 \!-\! \frac{1}{2} \ln(xy) \right] \!.
\ee
Due to the symmetry of the original triangle diagram with respect to all
the external legs, the following combinations are totally symmetric
in $p_1^2, p_2^2, p_3^2$:
\be
(p_3^2)^{-1} \; \Phi^{(1)}(x, y)
\hspace{15mm}
\mbox{and}
\hspace{15mm}
(p_3^2)^{-1} \; \left(
\Psi^{(1)}(x, y) + \textstyle{1\over3} \; \ln(xy) \; \Phi^{(1)}(x, y)
\right) .
\ee

The integrals (\ref{Phiint})--(\ref{Psiint})
can be evaluated in terms of polylogarithms,
and the results can be found in \cite{UD3} (eqs.~(11) and (29),
respectively). The results for $\Phi^{(1)}$ were also
presented in \cite{JPA,DT,UD1-2-4}.
Similar results for the triangle function in four dimensions
(involving dilogarithms) were presented in \cite{triangle},
while the $\ep$-part of massive triangle diagrams was considered in
\cite{NMB}.
Here, we present the results for $\Phi^{(1)}$
and $\Psi^{(1)}$ in a different form:
\bea
\label{Phi}
\Phi^{(1)}(x,y) = \frac{1}{2 \lambda} \left\{ \vphantom{1^1_1}
         4 \Li{2}{1-z_1} + 4 \Li{2}{1-z_2} + 4 \Li{2}{1-z_3}
\hspace{20mm} \right. \nonumber \\  \left.
         + \ln^2 z_1  + \ln^2 z_2 + \ln^2 z_3
         + 2 \ln x \ln z_1 + 2 \ln y \ln z_2 \right\} ,
\eea
\bea
\label{Psi}
{} \nonumber \\[-8mm]
\Psi^{(1)}(x,y) = -\frac{1}{2 \lambda}
\left\{ 4 \Li{3}{1-z_1^{-1}}
      + 4 \Li{3}{1-z_2^{-1}}
      + 4 \Li{3}{1-z_3^{-1}}
\right. \hspace{10mm} \nonumber \\[1mm]
      - 4 \Li{3}{1-z_1}
      - 4 \Li{3}{1-z_2}
      - 4 \Li{3}{1-z_3}
\hspace{17mm} \nonumber \\[1mm]
+ 4 \ln x \, \Li{2}{1-z_1}
+ 4 \ln y \, \Li{2}{1-z_2}
\hspace{30mm} \nonumber \\[1mm]  \left.
      - \ln z_1 \ln z_2 \ln z_3
      + \ln x \ln z_1 \ln (x z_1)
      + \ln y \ln z_2 \ln (y z_2)
 \vphantom{1^1_1} \right\} ,
\eea
where the variables $z_i$ and $\lambda$ are defined by
\be
z_1 = \frac{{(\lambda + x - y - 1)}^2}{4y}, \; \; \;
z_2 = \frac{{(\lambda + y - 1 - x)}^2}{4x}, \; \; \;
z_3 = \frac{{(\lambda + 1 - x - y)}^2}{4xy},
\ee
\be
\label{lambda}
\lambda(x,y) = \sqrt{(1-x-y)^2 - 4 x y}.
\ee
The $z_i$ are related to each other by $z_1 z_2 z_3 = 1$. The
representations~(\ref{Phi}) and (\ref{Psi}) are equivalent to the
ones presented in \cite{UD3} in the region $x+y<1$, as can be shown
by transformations of the trilogarithms and dilogarithms \cite{Lewin}.
They are also valid in the regions $x>y+1$ and $y>x+1$. When $\lambda$
approaches zero, the $z_i$ go to one, and hence all terms inside
the braces vanish\footnote{
In \cite{UD3}, the arguments of the  $\mbox{Li}_2$'s and $\mbox{Li}_3$'s
were chosen so as to vanish as $x$ and $y$ approach zero.}.
The analytic continuations of (\ref{Phi}) and (\ref{Psi})
into the region where $x+y>1$ and $x-1<y<x+1$, are obtained by adding
some logarithmic terms to them, in such a way that the discontinuities
along the branch cuts $z_i<0$ are cancelled.
Using the formulae of \cite{Lewin},
the resulting expressions can be written as follows\footnote{
Eq.~(\ref{Phi-Ls2}) corresponds to (4.15) of \cite{DT}.
Representations similar to (\ref{Phi-Ls2}) were also considered
in \cite{FJJ,Lu}.}:
\be
\label{Phi-Ls2}
\Phi^{(1)}(x,y)
= \frac{2}{\sqrt{-\lambda^2}}
\left\{ \Cl{2}{\theta_1}\!+\!\Cl{2}{\theta_2}\!+\!\Cl{2}{\theta_3} \right\}
= \frac{2}{\sqrt{-\lambda^2}}
\left\{ \Ls{2}{\theta_1}\!+\!\Ls{2}{\theta_2}\!+\!\Ls{2}{\theta_3} \right\},
\ee
\be
\label{Psi-Ls3}
\Psi^{(1)}(x,y)
= \ln\left(\frac{-\lambda^2}{xy}\right) \Phi^{(1)}(x,y)
- \frac{2}{\sqrt{-\lambda^2}}
\left\{ \Ls{3}{\theta_1} + \Ls{3}{\theta_2} + \Ls{3}{\theta_3}
+ \frac{\pi^3}{6} \right\} ,
\ee
\be
\label{theta-def}
\theta_1 = 2 \arccos\left(\frac{1\!-\!x\!+\!y}{2\sqrt{y}}\right) , \;\;
\theta_2 = 2 \arccos\left(\frac{1\!+\!x\!-\!y}{2\sqrt{x}}\right) , \;\;
\theta_3 = 2 \arccos\left(\frac{-1\!+\!x\!+\!y}{2\sqrt{xy}}\right) .
\ee
Note that $\theta_1+\theta_2+\theta_3=2\pi$.
The log-sine integral is defined by
\be
\Ls{N}{\theta} = - \Int_0^{\theta} \mbox{d} \theta \;
\ln^{N-1} \left| 2 \sin\frac{\theta}{2} \right| ,
\ee
and in particular, $\Ls{2}{\theta}\!=\!\Cl{2}{\theta}$.
Eqs.~(\ref{Phi-Ls2})-(\ref{theta-def}) are valid for all
$x, \; y$
such that $\lambda^2\!<\!0$.

In the problem under consideration, the parabola defined by
$\lambda(x,y)=0$ is a special curve (see also in \cite{FJJ}).
It consists of the three segments
$\sqrt{x} + \sqrt{y}=1$, $\; \sqrt{x} - \sqrt{y}=1$ and $\sqrt{y} -
\sqrt{x}=1$. On this curve, a simple result for arbitrary space-time
dimension can be obtained using eq.~(\ref{int-dr}),
\be
\label{J:lambda=0;xy}
\left. \!\! J(4-2\ep;1,1,1) \right|_{\lambda=0}
=  \frac{\pi^{2-\ep} \mbox{i}^{1+2\ep}}{(p_3^2)^{1+\ep}}
                 \frac{\Gamma (\ep) \Gamma^2 (1-\ep)}{\Gamma(2-2\ep)}
\left\{ \frac{x\!+\!y\!-\!1}{2xy}
                +  \frac{y\!+\!1\!-\!x}{2y}  x^{-\ep}
                +  \frac{1\!+\!x\!-\!y}{2x}  y^{-\ep} \right\} \!,
\ee
which is clearly symmetric in $p_1^2$, $p_2^2$ and $p_3^2$
and can be written as
\be
\label{J:lambda=0;p}
 - \frac{\pi^{2-\ep}\;\mbox{i}^{1+2\ep}}{p_1^2 \; p_2^2 \; p_3^2} \;
                 \frac{\Gamma (\ep) \Gamma^2 (1-\ep)}{\Gamma(2-2\ep)} \;
 \left\{ (p_1 p_2) (p_3^2)^{1-\ep}
                +(p_3 p_1) (p_2^2)^{1-\ep}
                +(p_2 p_3) (p_1^2)^{1-\ep} \right\} \; .
\ee
Expanding (\ref{J:lambda=0;xy}) in $\ep$ gives
\be
\label{PhiPsi|lambda=0}
\left. \!\! \Phi^{(1)}(x,y) \right|_{\lambda=0}
= - \frac{2 \ln x}{1\!-\!x\!+\!y}
          - \frac{2 \ln y}{1\!+\!x\!-\!y}  ; \; \;
\left. \Psi^{(1)}(x,y) \right|_{\lambda=0}
= \frac{ \ln^2 x \!-\! 4 \ln x}{1-x+y}
+ \frac{ \ln^2 y \!-\! 4 \ln y}{1+x-y} .
\ee

\vspace{4mm}

{\bf 4.}
Looking at the Mellin--Barnes representations (\ref{MBI})--(\ref{MBJ}),
it is possible to observe that
\bea
\label{connection1}
I(n; \nu_1, \nu_2, \nu_3) = \pi^{3n/2 - \Sigma\nu_i} \; \mbox{i}^{1-n}
\; \frac{\Gamma(\nu_2 + \nu_3 - n/2) \Gamma(\nu_1 + \nu_3 - n/2)
          \Gamma(\nu_1 + \nu_2 - n/2)}
        {\Gamma(\nu_1) \Gamma(\nu_2) \Gamma(\nu_3)}
\nonumber \\[2mm]
\times J\left( 2\Sigma\nu_i -n; \nu_2 + \nu_3 - n/2, \nu_1 + \nu_3 - n/2,
                          \nu_1 + \nu_2 - n/2 \right)  .
\eea
The same relation can be obtained from the
representations (\ref{FpI}) and (\ref{FpJ}) by first inverting
and then rescaling the Feynman parameters
$\alpha_i$ ($i=1,2,3$) in (\ref{FpI}):
\be
\alpha_i=\frac{1}{\alpha'_i} , \; \; \; \;
\alpha'_i= {\cal{F}}(\alpha''_1,\alpha''_2,\alpha''_3) \alpha''_i  , \; \;
\; \;
{\cal{F}}(\alpha''_1,\alpha''_2,\alpha''_3)
    = \frac{\alpha''^{-1}_1+\alpha''^{-1}_2+\alpha''^{-1}_3}
                   {\alpha''_1+\alpha''_2+\alpha''_3}  ,
\ee
where the scaling factor ${\cal{F}}$ has been chosen to restore the argument
of the delta function to its original form. Due to the homogeneity
of the integrand, the effect of the rescaling is to multiply it
by a factor ${\cal{F}}^{-3}$, which is precisely cancelled by the Jacobian
associated with the change of variables $\alpha'_i \to \alpha''_i$
\cite{Scharf+ST}. The result has the structure of (\ref{FpJ}).

One's first impression may be that the relation (\ref{connection1}) does
not look very useful. In particular, in the case
$n=4-2\ep, \; \nu_1=\nu_2=\nu_3=1$ it yields
\be
I(4-2\ep; 1, 1, 1) = \pi^{3-3\ep} \; \mbox{i}^{1+2\ep}
\; \Gamma^3(\ep) \; J(2+2\ep; \ep, \ep, \ep) ,
\ee
with some ``strange'' integral on the r.h.s. However, we can use
the uniqueness relation (\ref{uniq}) to transform the integrals $J$.
Applying (\ref{uniq}) with respect to all three external legs of
the three-point diagram gives:
\bea
\label{uniq2}
\Gamma(\nu_1) \Gamma(\nu_2) \Gamma(\nu_3) \Gamma(n- \Sigma\nu_i)
\; J\left( n; \nu_1, \nu_2, \nu_3 \right)
\hspace{75mm}
\nonumber \\[1mm]
= \Gamma(n/2 -\nu_1) \Gamma(n/2 -\nu_2) \Gamma(n/2 -\nu_3)
  \Gamma(\Sigma\nu_i -n/2)
\hspace{50mm}
\nonumber \\[1mm]
\times (p_1^2)^{n/2 - \nu_2 - \nu_3}
       (p_2^2)^{n/2 - \nu_1 - \nu_3}
       (p_3^2)^{n/2 - \nu_1 - \nu_2}
       \; J\left( n; n/2 -\nu_1, n/2 -\nu_2, n/2 -\nu_3 \right) .
\eea

Combining (\ref{connection1}) and (\ref{uniq2}), we get
\be
\label{connection2}
I(n; \nu_1, \nu_2, \nu_3) = \pi^{3n/2 - \Sigma\nu_i} \; \mbox{i}^{1-n}
\left( \prod_{i=1}^{3} (m_i^2)^{n/2 -\nu_i} \right)
\frac{\Gamma(\sum \nu_i \!-\!n)}{\Gamma(n/2)} \;
J\left( 2\Sigma\nu_i\! -\!n; \nu_1, \nu_2, \nu_3 \right) .
\ee
Now, the powers $\nu_i$ are the same on the l.h.s. and on the r.h.s.
whilst the values of the space-time dimension are different.
We shall call eq.~(\ref{connection2}) a ``magic'' connection.
It can also be derived by a change of variables in the Feynman
parametric representation or in the Mellin--Barnes contour integrals.

Let us consider what the ``magic'' connection gives for the most
interesting case $\nu_1 = \nu_2 = \nu_3 = 1$ at different values of $n$.

For $n=2-2\ep$, we get
\bea
\label{I2-J4}
{} \nonumber \\[-9mm]
I(2-2\ep; 1, 1, 1) = \pi^{-3\ep} \; \mbox{i}^{-1+2\ep}
\; (m_1^2 m_2^2 m_3^2)^{-\ep}
\; \frac{\Gamma(1+2\ep)}{\Gamma(1-\ep)} \;
J(4+2\ep; 1, 1, 1) .
\eea
So, the dimensionally-regularized triangle integral $J(4+2\ep; 1, 1, 1)$
can be related to the two-dimensional integral $I(2-2\ep; 1, 1, 1)$
(also dimensionally-regularized, but note that signs of $\ep$ are
different!). Since both integrals are convergent as $\ep \to 0$,
we can put $\ep=0$:
\be
I(2; 1, 1, 1) = - \mbox{i} \; J(4; 1, 1, 1)
= \pi^2 \; m_3^{-2} \; \Phi^{(1)}(x,y) .
\ee

For $n=3-2\ep$, we get
\bea
\label{I3-J3}
{} \nonumber \\[-8mm]
I(3-2\ep; 1, 1, 1) = \pi^{3/2-3\ep} \; \mbox{i}^{2+2\ep}
\; (m_1^2 m_2^2 m_3^2)^{1/2 -\ep}
\; \frac{\Gamma(2\ep)}{\Gamma(3/2 -\ep)} \;
J(3+2\ep; 1, 1, 1) .
\eea
In this case, both integrals are three-dimensional, but (again!) the
signs of $\ep$ are different on the l.h.s. and on the r.h.s.
Note that on the r.h.s. we also have a singular factor $\Gamma(2\ep)$.
So, to get the result for the singular and ``constant''
(in $\ep$) terms of $I(3-2\ep; 1, 1, 1)$, we need to know
$J(3+2\ep; 1, 1, 1)$ up to the $\ep$-part.
Using the representation (\ref{int-dr}), it can be
easily calculated, and we arrive at the following result:
\be
I(3-2\ep; 1, 1, 1) = \pi^{4-2\ep} \; \Gamma^2(1+\ep) \;
\left\{ \frac{1}{\ep} + 2 - 4 \ln\left(m_1 + m_2 + m_3\right) \right\}
+ {\cal{O}}(\ep) .
\ee
This corresponds to the result presented in
\cite{Shaposh}, eq.~(110).
Note that simple results for three-dimensional triangles were
presented in \cite{triangle3}.

For $n=4-2\ep$,
eq.~(\ref{connection2}) gives:
\bea
{} \nonumber \\[-7mm]
I(4-2\ep; 1, 1, 1)
 = \pi^{3-3\ep} \; \mbox{i}^{1+2\ep}
\; (m_1^2 m_2^2 m_3^2)^{1-\ep}
\; \frac{\Gamma(-1+2\ep)}{\Gamma(2-\ep)} \;
J(2+2\ep; 1, 1, 1) .
\eea
Using Feynman parameters, it is easy to show that
\be
\!\! J\left( 2+2\ep; 1, 1, 1 \right)
= - \pi^{-1}  \left\{ J\left( 4+2\ep; 2, 1, 1 \right)
                        \!+\! J\left( 4+2\ep; 1, 2, 1 \right)
                        \!+\! J\left( 4+2\ep; 1, 1, 2 \right) \right\}.
\ee
Now, we use the formula \cite{JPA} obtained by integration by parts
\cite{ibp},
\bea
\label{J112}
J(4+2\ep; 1, 1, 2) =
(p_1^2 \; p_2^2)^{-1} \;
  \left \{ - (p_1^2  + p_2^2  - p_3^2 ) \;
             \varepsilon  J(4+2\ep; 1, 1, 1) \right.
\hspace{30mm}
\nonumber \\
\left.
         + p_1^2  J(4+2\ep; 0, 2, 1)
         + p_2^2  J(4+2\ep; 2, 0, 1)
         - p_3^2  J(4+2\ep; 2, 1, 0)
\right \}
\eea
(note that the sign of $\ep$ is different than in \cite{JPA}).
The integrals $J$ with one of the $\nu$'s equal to zero correspond
to massless one-loop two-point functions.
In such a way, we get
\bea
\left\{ J\left( 4+2\ep; 2, 1, 1 \right)
      + J\left( 4+2\ep; 1, 2, 1 \right)
      + J\left( 4+2\ep; 1, 1, 2 \right) \right\}
\hspace{45mm}
\nonumber \\[1mm]
= (p_1^2 p_2^2 p_3^2)^{-1} \;
\left\{ - \ep \Delta(p_1^2, p_2^2, p_3^2)  \; J(4+2\ep; 1, 1, 1)
+ p_1^2  (-p_1^2 + p_2^2 + p_3^2) J(4+2\ep; 0, 2, 1)
\right.
\nonumber \\[1mm]
\left.
+ p_2^2  (p_1^2 - p_2^2 + p_3^2) J(4+2\ep; 2, 0, 1)
+ p_3^2  (p_1^2 + p_2^2 - p_3^2) J(4+2\ep; 2, 1, 0)
\right\}
\eea
where
\bea
\label{Kallen}
{} \nonumber \\[-8mm]
\Delta(p_1^2, p_2^2, p_3^2)
=  2 p_1^2 p_2^2 + 2 p_1^2 p_3^2 + 2 p_2^2 p_3^2
 - (p_1^2)^2 -  (p_2^2)^2 -  (p_3^2)^2
= -(p_3^2)^2 \; \lambda^2(x,y)
\eea
is connected with the K\"all\'en function.

Thus, the connection can be written in the following symmetric form:
\bea
\label{sym_form}
I(4-2\ep; 1, 1, 1) = - \textstyle{1\over2} \; \pi^{4-2\ep} \;
(m_1^2 m_2^2 m_3^2)^{-\ep}
\; \left[ (1-\ep) (1-2\ep) \right]^{-1}
\hspace{42mm}
\nonumber \\[1mm]
\times\! \left\{ - \frac{1}{\pi^{2+\ep} \mbox{i}^{1-2\ep}} \;
\frac{\Gamma(1+2\ep)}{\Gamma(1-\ep)} \;
J(4+2\ep; 1, 1, 1) \; \Delta(m_1^2, m_2^2, m_3^2)
\right.
\hspace{45mm}
\nonumber \\
+ \Gamma^2(\ep)
\left[ (m_1^2)^{\ep} (-m_1^2 \!+\! m_2^2 \!+\! m_3^2)
      + (m_2^2)^{\ep} (m_1^2 \!-\! m_2^2 \!+\! m_3^2)
\left. \frac{}{}
      + (m_3^2)^{\ep} (m_1^2 \!+\! m_2^2 \!-\! m_3^2)
\right]
\right\}\! .
\eea

Using (\ref{J-ep}) to expand in $\ep$ and keeping the terms up to
the order $\ep$, we get
\bea
\label{I-ep}
I(4-2\ep; 1, 1, 1) = \textstyle{1\over2} \; \pi^{4-2\ep} (m_3^2)^{1-2\ep}
\; \Gamma^2(1+\ep) \left[ (1-\ep) (1-2\ep) \right]^{-1}
\hspace{35mm}
\nonumber \\[1mm]
\times \left\{ - \frac{1}{\ep^2} \; (1+x+y)
+ \frac{2}{\ep} \; (x \ln x + y \ln y)
\right.
- x \ln^2 x - y \ln^2 y + (1-x-y) \ln x \ln y
\nonumber \\
+ \textstyle{1\over3} \ep
(x \ln^3 x + y \ln^3 y)
- \textstyle{1\over2} \ep (1-x-y) \ln x \; \ln y \;(\ln x + \ln y)
\hspace{25mm}
\nonumber \\[1mm]
\left. \frac{}{}
- \lambda^2(x, y) \left( 1 - \ep (\ln x + \ln y) \right)
\; \Phi^{(1)}(x,y)
+ \ep \; \lambda^2(x, y) \; \Psi^{(1)}(x, y)
\right\} + {\cal O}(\ep^2) \; .
\hspace{3mm}
\eea
The divergent and constant (in $\ep$) terms coincide with
the result of \cite{DT}, eq.~(4.9).

It could be noted that, using eqs.~(\ref{sym_form}) and (\ref{I2-J4}),
we can write an exact relation between integrals $I$ with
different values of the space-time dimension:
\bea
\label{I4-I2}
I(4-2\ep; 1, 1, 1) = - \textstyle{1\over2} \pi^{4-2\ep}
\left[(1-\ep) (1-2\ep) \right]^{-1}
\left\{ - \pi^{-2+2\ep}
 \Delta(m_1^2, m_2^2, m_3^2) I(2-2\ep; 1, 1, 1)
\right.
\hspace{-4mm}
\nonumber \\[1mm]
\left.
+ \Gamma^2(\ep) (m_1^2 m_2^2 m_3^2)^{-\ep}
\!\left[ (m_1^2)^{\ep} (-m_1^2 \!+\! m_2^2 \!+\! m_3^2)
      \!+\! (m_2^2)^{\ep} (m_1^2 \!-\! m_2^2 \!+\! m_3^2)
      \!+\! (m_3^2)^{\ep} (m_1^2 \!+\! m_2^2\! -\! m_3^2)
\right]
\right\} \! .
\hspace{-10mm}
\nonumber \\
{}
\eea
Using
(\ref{connection2}), an analogous relation
can also be written for one-loop triangles, namely:
\bea
\label{J2-J4}
J(2+2\ep; 1,1,1) =
(p_1^2 p_2^2 p_3^2)^{-1} \;
\left\{
\pi^{-1} \;
\ep \; \Delta(p_1^2, p_2^2, p_3^2) \;
J(4+2\ep; 1,1,1)
\right.
\hspace{32mm}
\nonumber \\[2mm]
\left.
+ 2 \pi^{1+\ep} \;{\mbox{i}}^{1-2\ep} \;
\frac{\Gamma(1-\ep) \Gamma^2(1+\ep)}{\Gamma(1+2\ep)} \;
\frac{1}{\ep} \;
\left[ (p_1 p_2)(p_3^2)^{\ep} + (p_3 p_1)(p_2^2)^{\ep}
       + (p_2 p_3)(p_1^2)^{\ep} \right]
\right\} .
\eea
This is a special case ($N\!=\!3$) of an identity given in \cite{BDK}
relating one-loop $N$-point integrals to $(N\!-\!1)$-point ones.
Note that at $\lambda=0$ ($\Delta=0$) the first term in the braces
on the r.h.s. disappears, and, changing $\ep$ into $(1-\ep)$, we
obtain nothing but the result (\ref{J:lambda=0;p}).

\vspace{4mm}

{\bf 5.}
In many realistic applications, the masses of internal particles
are equal.
For this case ($m_1=m_2=m_3 \equiv m$, \ $x=y=1$),
the integral representations (\ref{Phiint}) and (\ref{Psiint}) give
\be
\label{phipsi111int}
\Phi^{(1)}(1,1)= - 2 \Int_0^1
\frac{\mbox{d} \xi}{1 \!-\!\xi \!+\! \xi^2} \;
\ln\xi , \; \; \; \;
\Psi^{(1)}(1,1)= - 2 \Int_0^1
\frac{\mbox{d} \xi}{1 \!-\! \xi \!+\! \xi^2} \;
\ln\xi \; \ln\left( \frac{1 \!-\!\xi \!+\! \xi^2}{\xi} \right).
\ee
We note that the first of these two integrals was
also presented in \cite{CelGon}.

The angles (\ref{theta-def}) are now all equal to $2\pi/3$,
so that eqs. (\ref{Psi-Ls3}) and (\ref{Phi-Ls2}) are reduced
to\footnote{
The representation in terms of $\Cl{2}{\textstyle{{\pi}\over3}}$
is well-known (see e.g. in \cite{Bij,Bro'90}).
$\Phi^{(1)}(1,1)$
can be also expressed as
$\Phi^{(1)}(1,1) =
\frac{2}{3} \left[ \psi'\left(\frac{1}{3}\right)
- \frac{2}{3} \pi^2 \right] $, see in \cite{Canada}.}
\be
\label{Phi11-Cl}
\Phi^{(1)}(1,1)
= \textstyle{6\over{\sqrt{3}}} \; \Cl{2}{\textstyle{{2\pi}\over3} }
= \textstyle{4\over{\sqrt{3}}} \; \Cl{2}{\textstyle{{\pi}\over3}}  ,
\ee
\be
\label{Phi11-Ls}
\Psi^{(1)}(1,1)= \textstyle{1\over{\sqrt{3}}} \;
\left\{ -6 \; \Ls{3}{\textstyle{{2\pi}\over3}}
       + 4 \; \ln{3} \; \Cl{2}{\textstyle{{\pi}\over3}}
        - \textstyle{1\over3} \; \pi^3
\right\} .
\ee
Eqs.~(\ref{Phi11-Cl})-(\ref{Phi11-Ls}) can be expressed
in terms of the generalized inverse tangent integral \cite{Lewin},
\be
\Ti{N}{z}
= \frac{1}{2 \; \mbox{i}}
\left(\Li{N}{\mbox{i} z} - \Li{N}{-\mbox{i} z} \right)
= \frac{(-1)^{N-1}}{(N-1)!} \; z
\Int_0^1 \mbox{d} \xi \frac{\ln^{N-1}\xi}{1+z^2 \xi^2} \; ,
\ee
whose Taylor series is
\be
\Ti{N}{z} = z \sum\limits_{j=0}^{\infty} \frac{(-z^2)^j}{(2j+1)^N}
= \frac{z}{1^N} - \frac{z^3}{3^N} + \frac{z^5}{5^N} - \ldots \; ,
\hspace{10mm} |z| < 1 \; .
\ee
Using
\be
\Ti{3}{\textstyle{1\over{\sqrt{3}}}} = {\textstyle{1\over{\sqrt{3}}}}
\; \sum\limits_{j=0}^{\infty} \frac{(-1)^j}{3^j \; (2j+1)^3}
= 0.570681635...
\ee
\be
\Cl{2}{\textstyle{{\pi}\over3}} = \textstyle{6\over5} \;
            \Ti{2}{\textstyle{1\over{\sqrt{3}}}}
                        + \textstyle{1\over{10}} \; \pi \; \ln{3}
=1.014941606... ,
\ee
\be
\Ls{3}{\textstyle{{2\pi}\over3}} =
  \textstyle{8\over5} \; \Ti{3}{\textstyle{1\over{\sqrt{3}}}}
 + \textstyle{2\over3} \; \ln{3} \; \Cl{2}{\textstyle{{\pi}\over3}}
 - \textstyle{1\over{30}} \; \pi \; \ln^2{3}
 - \textstyle{{16}\over{135}} \; \pi^3
 = -2.144767213...
\ee
we find
\be
\Phi^{(1)}(1,1)
= \textstyle{2\over{5 \sqrt{3}}}
  \left\{ 12 \; \Ti{2}{\textstyle{1\over{\sqrt{3}}}}
                              + \pi \; \ln{3} \right\}
 = 2.343907239... ,
\ee
\be
\Psi^{(1)}(1,1)= \textstyle{1\over{5 \sqrt{3}}} \;
\left\{ - 48 \; \Ti{3}{ \textstyle{ 1\over{\sqrt{3}} } }
        +  \pi \; \ln^2{3} + \textstyle{17\over9} \; \pi^3 \right\}
 = 4.037576132... \,.
\ee

\vspace{4mm}



{\bf 6.}
In this paper we have derived a useful relation (\ref{connection2})
between two {\em very different} types of Feynman diagrams:
two-loop massive vacuum integrals (\ref{defI}) and
one-loop three-point functions (\ref{defJ}).
This ``magic'' connection is valid for any values of the
space-time dimension and the powers of propagators.
While the powers of propagators are the same,
the massive and massless integrals related by (\ref{connection2})
have different values of the space-time dimension.
Nevertheless, using some additional transformations it
is possible to relate dimensionally-regulated integrals
considered around $n=4$ (see eq.~(\ref{sym_form})).
However, we get different signs of $\ep$ on the l.h.s.
and on the r.h.s. ($n=4\mp 2\ep$). As a result, some
ultraviolet singularities of massive integrals can correspond
to infrared singularities of massless triangles.
This is not dangerous, since the magic connection relates
diagrams of different type and does not introduce additional
mixing of different singularities
in diagrams of the same type.

An important application considered in the paper is the result
for the $\ep$-part of the two-loop massive vacuum diagram with different
masses, eq.~(\ref{I-ep}). For the equal-mass case, a new transcendental
constant is shown to appear. It is not excluded that it can be
connected with an analytically-unknown constant occurring in
three-loop calculations of the $\rho$-parameter \cite{3-loop-rho}
(see also in \cite{Bro'92}).

Using integral representations (\ref{int-dr})--(\ref{secret}),
higher terms of the expansion in $\ep$ (and in $\delta$)
can also be obtained. Moreover, the corresponding non-trivial
functions will be the same in both cases, due to the
magic connection.

\vspace{2mm}

{\bf Acknowledgements.}
We are grateful to  F.A.~Berends, D.J.~Broadhurst, P.~Osland,
M.E.~Shaposhnikov and O.V.~Tarasov for useful discussions. J.B.T. thanks the
University of Bergen, and A.D. the University of Mainz,
where parts of this work have been done, for hospitality.
A.D.'s research was supported by the Research Council of Norway.
J.B.T. was supported by the Graduiertenkolleg
``Teilchenphysik'' in Mainz.

\vspace{4mm}


\end{document}